\documentclass[11pt,twoside]{article}
\usepackage{./asp2014}

\aspSuppressVolSlug
\resetcounters

\begin{document}

\title{Triple Stars Observed by {\em Kepler}}
\author{Jerome A. Orosz,$^1$}
\affil{$^1$Department of Astronomy, San Diego State University,
San Diego, California, United States; \email{jorosz@mail.sdsu.edu}}

\paperauthor{Jerome A. Orosz}{jorosz@mail.sdsu.edu}{}{San Diego State University}{Department of Astronomy}{San Diego}{CA}{92182-1221}{United States}

\begin{abstract}
The Kepler mission has provided high quality light curves for more
than 2000 eclipsing binaries. Tertiary companions to these binaries
can be detected if they transit one or both stars in the binary or if
they perturb the binary enough to cause deviations in the observed
times of the primary and secondary eclipses (in a few cases both
effects are observed in the same eclipsing binary). From the study of
eclipse timing variations, it is estimated that 15 to 20\% of the
Kepler eclipsing binaries have close-in tertiary companions.  I will
give an overview of recent results and discuss some specific
systems of interest.
\end{abstract}

\section{Introduction}
In an isolated, detached eclipsing binary (EB), the eclipses
should be strictly periodic, with a constant
interval of time between successive
eclipses.   In these cases, the times of eclipse are
described by a simple linear ephemeris:
\begin{equation}
T_{\rm min}(E)=T_0+P_{\rm bin}E
\end{equation}
where $P_{\rm bin}$ is the binary orbital period and $E$ is the
cycle number.  The residuals derived from a fit to a linear
ephemeris form an ``Observed minus Computed'' or O-C diagram.
If one measures eclipse times (ETs) for both primary and secondary eclipses,
then both types of events may be put on a common system by means of
a phase offset $\delta$ that is applied to the cycle numbers of the
secondary eclipses.
A single period can then
be fit to all
of the ETs.
We will refer to the resulting O-C
diagram a ``Common Period O-C'' or CPOC diagram.

There are a number of situations where the intervals
between successive eclipses
are  not constant, thereby leading to
eclipse timing variations (ETVs).  The points in the O-C or CPOC diagrams
will no longer be scattered about the horizontal axis, and a model that
goes beyond a simple linear ephemeris will be needed.  We briefly discuss
mechanisms for ETVs that apply to stars that are well within their 
respective Roche lobes, where mass transfer and mass loss can be neglected.

If the EB is part of a triple system, 
then the eclipses will either be early or late owing to light travel
time (LTT) changes as the EB moves about the center of mass of the triple
system.  In these cases, the ETs are
no longer described by a simple linear ephemeris (e.g.\ Irwin 1952):
\begin{equation}
T_{\rm min}(E)=T_0+P_{\rm bin}E
+K\left[{1-e_3^2\over 1+e_3\cos\nu_3}\sin(\nu_3+\omega_3)+e_3\sin\omega_3\right],
\label{ocorbit}
\end{equation}
where $e_3$ is the eccentricity and  $\omega_3$ is the longitude of periastron
of the tertiary orbit, 
$\nu_3$ is the true anomaly in that orbit,
and where the semi-amplitude $K$ is given by
\begin{equation}
K={a_{12}\sin i^3\over c},
\end{equation}
where $a_{12}$ is the semimajor axis
of the EB orbit about the third star, $i_3$ is the inclination to
the observer's line-of-sight, and where $c$ is the speed of light.
The minimum mass of the third body can be computed in a way that
is similar to the well-known result that can be derived from
a radial velocity curve:
\begin{equation}
f(M_3)={M^3_3\sin^3i_3\over
(M_1+M_2+M_3)^2}={4\pi^2(a_{12}\sin i_3)^3\over GP^2_3}
\label{fm}
\end{equation}
where $M_1$ and $M_2$ are the masses of the stars in the EB,
$M_3$ is the tertiary mass, and $G$ is the constant of
gravitation. 
Many close EBs are found to have third components based on LTT studies
(e.g.\ Tokovinin et al.\ 2006).  

If the third star is in a relatively close orbit with the EB, then
the
changing proximity of the third body can affect the orbital period
of
the inner binary.  
Borkovits et al.\ (2003)
presented an analytic study of the ETVs
caused by an
exterior, perturbing object on a large, eccentric orbit.  The 
signals one sees in the O-C diagrams can be quite complex, depending on the
eccentricities of the EB and outer orbit, and on the mutual inclinations of the
orbits.  

Before the launch of {\em Kepler}, EBs where
dynamical effects dominate the ETVs
were comparatively rare.  One example is
IU Aurigae (\"Ozdemir et al.\ 2003), which is a massive 
binary ($M_1=21.4\,M_{\odot}$ and $M_2=14.5
\,M_{\odot}$) with an orbital period of 1.8 days.  There is 
a third body with a period of 293 days which causes precession
of the inner binary
on a time-scale of $\approx 335$ years.  If the
ETVs are taken to be from LTT effects
alone, then
the minimum mass of the third body is about $15\,M_{\odot}$,
which is much too massive, given the amount of third light
measured from spectroscopy and from the light curve solutions.
Hence \"Ozdemir et al.\ (2003) argue that a large part of the signal seen
in the O-C diagram is due to dynamical effects.
Another example is
SS Lac, which was known to be a deeply eclipsing EB around the year 1900, but
then stopped eclipsing around the year 1950, presumably due to the
influence of a third body.  This   
outer body with
a period of 697 days was detected spectroscopically by
Torres \& Stefanik (2000).  In a subsequent paper, Torres (2001)
modeled the depth changes of the eclipses and found a precession period
of about 600 years.  More recently, 
Graczyk et al.\ (2011) found 17 EBs in the LMC via the OGLE survey
where the eclipse depths changed.  In two cases the eclipses disappeared
altogether.


The line of apsides of an eccentric binary can rotate due to
the effects of General Relativity. The ratio of
the orbital period of the binary to the rotation period $U_{\rm GR}$
of the line of apsides is given by
\begin{equation}
{P_{\rm bin}\over U_{\rm GR}}= 6.35\times 10^{-6}
\left({M_1 + M_2\over a(1-e^2)}\right)
\end{equation}
where $M_1$ and $M_2$ are the masses of each star in solar masses, and
$a$ is the semimajor axis of the relative orbit in solar radii
(Kopal 1978).
In addition, 
tidal distortions can cause the line of apsides of an eccentric binary
to rotate with a period $U_{\rm tide}$ given by 
\begin{equation}
{P_{\rm bin}\over U_{\rm tide}}=c_{21}k_{21}+c_{22}k_{22}\label{apeq1}
\end{equation} 
where $k_{12}$ and $k_{22}$ are the so-called internal
structure constants of the stars, and where
the coefficients $c_{21}$ and $c_{22}$ are given by
\begin{eqnarray}
c_{21} & = & \left[\left({\omega_1\over\omega_K}\right)^2
\left(1+{M_2\over M_1}\right)f(e)+{15M_2\over M_1}g(e)\right]
\left({R_1\over a}\right)^5\label{apeq2}  \nonumber \\
c_{22} & = & \left[\left({\omega_2\over\omega_K}\right)^2
\left(1+{M_1\over M_2}\right)f(e)+{15M_1\over M_2}g(e)\right]
\left({R_2\over a}\right)^5.\label{apeq3}
\end{eqnarray}
In the above equations,
($\omega_1/\omega_K$) and ($\omega_2/\omega_K$) are the ratios between
the actual angular rotational velocity of the stars and the
rotational velocity corresponding to synchronization with the
average orbital velocity.
The functions $f(e)$ and $g(e)$ are given by
\begin{eqnarray}
f(e)&=&{1\over (1-e^2)^2}\label{apeq4} \nonumber \\
g(e)&=&{(8+12e^2+e^4)f(e)^{2.5}\over 8}\label{apeq5}
\end{eqnarray}
(Claret \& Gim\'enez 1993).  Note that the expected apsidal period
depends on the fifth power of the fractional radii of the stars, and that
for EBs with periods more than about 15 to 20 days the apsidal period
will usually be factors of several million or more times the binary period,
assuming the EBs contain solar-type main sequence stars.

Since the phase difference between the primary
and secondary eclipses changes with time, 
``apsidal motion'' due to General Relativity or to tides
will give rise to ETVs, where the O-C curves of the primary and secondary 
roughly resemble sine curves that are 180 degrees out of phase.
Unfortunately there are no simple close-form
expressions to model the signal seen in the O-C diagram of an eccentric
EB undergoing apsidal motion. 
Gim\'enez \& Garcia-Pelayo (1983) give
a power series expression up to the fifth power of the
eccentricity that contains roughly 50 terms.  Lacy (1992)
presented an exact solution based on an iterative solution
of the transcendental equations involved.

\section{The {\em Kepler} EB Sample}
The {\em Kepler} mission 
\citep{Borucki_2010}
observed over 2000 EBs
with a duty cycle on the order of 90\% or greater, and good
photometric precision (Pr\v{s}a et al.\ 2011, Slawson et al.\ 2011).  
These data give us the unique opportunity
to measure precise ETs and detect deviations from
a linear ephemeris.   
Previously, 
Gies et al.\ (2012)
measured ETs for 41 {\em Kepler} EBs using data through
Q9 ($\approx 2$ year baseline).  They found no evidence for short-period
companions with periods smaller than $\approx 700$ days
in the sample, but did find evidence for long-term trends in
14 systems.  Rappaport et al.\ (2013) did a much more thorough survey
where they measured ETs for 2175 {\em Kepler} EBs using data through
Q13 ($\approx 3$ year baseline).  They identified 39 candidate
triple systems.   Conroy et al.\ (2014) presented a catalog
of ETs for 1279 close binaries (e.g.\ the overcontact and
ellipsoidal EBs with periods generally smaller than 1 day) in the
{\em Kepler} sample.  There were 236 systems where the ETVs
were compatible with the presence of a third body. 

We give an update on our own program to measure accurate 
ETs for the {\em Kepler} EBs with periods longer than about
1 day.  We have developed a suite of algorithms and codes
(collectively called ``TEMPUS'') 
to automatically measure times of mid eclipse for detached
{\em Kepler} EBs.  TEMPUS was developed on
and runs in the Matlab environment.
An earlier version of the software was
described in Steffen et al.\ (2011).  TEMPUS has
evolved since then, so a new overview is presented here.

TEMPUS is a computational system for measuring ETs
from the {\em  Kepler} SAP light curves.
Briefly, a model eclipse profile constructed from the data is used
to find individual ETs.  Because the ETs may not
be described by a simple linear ephemeris, an iterative method is used.
The iteration
begins with an initial set of ET values, usually 
computed from a linear ephemeris.
The eclipses in the data are found and locally detrended using a 
cubic spline.
The detrending algorithm
requires that each eclipse event has well-defined ``shoulders'' 
so that the first and fourth contact points can be identified.  
When such shoulders
are present, we have found the local detrending works well in most cases.
Once the individual eclipse events are locally detrended, they are
folded to produce an eclipse profile.
A piecewise cubic Hermite spline (PCHS)
model is then fit to this profile.  
During this
process, nearby data are tested for gaps
and other problems that would compromise
the determination of that ET
value.  Those ETs that exhibit such problems are eliminated from further
consideration.  The PCHS is then used to improve the ET estimates, and
those, in turn, provide an improved eclipse profile constructed
from the local detrending and folding.  
This
iteration is run three times.  At this point the ET estimates are
significantly improved and the eclipses with
compromised data  have been
eliminated.  However, there remains a chance that a
good  eclipse may have
been eliminated as the ET estimates were being adjusted. 
Therefore, using the latest ET
estimates this iterative process is restarted with any missing ET
estimates being approximated by nearby ones.  After three additional
iterations, the so-called Pipeline has produced very good ET estimates and a
very good PCHS model of the eclipse profile.

The final step in the  process is to compute an unbinned PCHS
model that when binned to the Kepler Long Cadence exposure
time (29.4244 minutes)
will best represent
the eclipse profile made from the detrended and folded data.  
As an initial approximation, the PCHS 
resulting from the six iterations is used.
It is binned for each point
in the latest locally-detrended and folded data and used to improve
the ET estimates.  These ET estimates again determined a revised local
detrending and folding of the data with a best, unbinned PCHS model
being computed and, in turn determining a further, but very minor,
revision in the ET estimates.  The last iteration of this process
occurs with the final result being a PCHS model that, when binned, 
optimally
describes the eclipse profile produced from the
locally detrended and folded data, which are determined by the
latest ET estimates.

During this process, ETs that are outliers to the individual model
fits are eliminated from defining further models but remain as values
to be estimated.  The error estimates for each ET value use the
standard approach of sliding the model, in time, across the ET
estimate and noting when the $\chi^2$ curve rises above the $1\sigma$
threshold.

Producing a fully automated process to measure 
ETs for {\em Kepler} EBs is a difficult task owing to the
wide variety of light curve morphologies (e.g.\ see Pr\v{s}a et
al.\ 2011 and Slawson et al.\ 2011):   Many binaries are spotted, many
of them have pulsations, many binaries have deep eclipses, and
others are severely diluted.  In order to have confidence in the
results, the TEMPUS pipeline
produces many diagnostic plots such as plots
of the individual O-C and CPOC diagrams, plots of the observed
and model eclipse profiles, power spectra of the ETVs, and
plots of a small subset of the ``raw'' light curve.  These
plots were inspected by eye and interesting systems were selected.

\section{Results}

We used all of the available long cadence data
through Q16 as input to TEMPUS.  
Our initial sample consisted of the 1322 binaries from the catalog
of Slawson et al.\ (2011) classified as either detached or
semidetatched, supplemented with some additional EBs that were
discovered since that publication.  
As the orbital period gets shorter and
shorter, reflection effects and
tidal distortion tend to make the 
out-of-eclipse regions increasingly  curved.  Since our
technique requires well-defined eclipse shoulders to work, we placed a cutoff
at orbital periods shorter than 0.9 days.  This cutoff removed a few
hundred  binaries
from the sample, leaving 1258.  
These binaries were fed into TEMPUS
to measure the ETs, 
and 1249 systems had their primary ETs measured
successfully.  The cases that failed either 
lacked well-defined shoulders, or had very
low signal-to-noise eclipses.  Eccentric 
binaries can have primary eclipses and
no secondary eclipses, so the sample of binaries 
with measurable secondary eclipses
is somewhat smaller.  In the end, a total of 764 systems had both primary and
secondary ETs measured successfully.  The typical uncertainties on
the ETs range from a few seconds to several minutes for the
noisy cases.

The diagnostic plots from TEMPUS
were inspected and obvious spurious cases were
removed.  Particular attention was paid to the
plots showing the final eclipse profile and final PCHS model.  
%
Based on the quality of the fits to the eclipse profiles, 335
EBs were removed from the sample, leaving 914 systems.

\articlefiguretwo{oroszfig01.ps}{oroszfig02.ps}{smallrms}{\emph{Left:} 
Example CPOC diagrams where the rms scatter in the ETVs is 
smaller than
$\approx 10$ sec.  The black points show the primary
ETVs and the red points show the secondary ETVs.
From top to bottom, the systems are
KIC 9913481, KIC 10156064, KIC 10191056, and KIC 12356914.
\emph{Right:} The distribution of the rms of the primary ETVs
in minutes. Most of the systems have reasonably small scatter, although
the tail of the distribution extends to much higher rms values.
}

Finally, for each binary, the O-C diagrams for the primary eclipses
and secondary eclipses (if present) were made.
In cases where both primary and
secondary eclipses were present, an 
iterative procedure was used to
produce a CPOC diagram.   These plots were visually
inspected, and interesting cases were selected.  We give below
an overview of these results.

\subsection{EBs With Small ETVs}

One simple statistic to compute is the rms of the primary ETVs.  A
large rms in the ETVs may indicate the presence of a 
nearby third body, although one needs to verify this on a case-by-case
basis (for example, star spots can induce spurious ETV signals as discussed
below).  
Likewise, a small rms in the ETVs might be used to place
limits on the lack of a third body, although, again, one needs to
verify each case individually.  There may be cases where long-term
trends are evident in spite of the small rms, or cases where
the primary and secondary ETV signals diverge in the CPOC.  
Figure \ref{smallrms} shows
four examples where the primary and secondary
rms values are smaller than about 10 seconds.  Figure 1
also shows the distribution of rms values for the primary ETVs.  
The histogram peaks around 10 seconds.  A total of 502 systems
have rms values smaller than 30 seconds and 741 systems have rms values
of smaller than one minute.

\articlefiguretwo{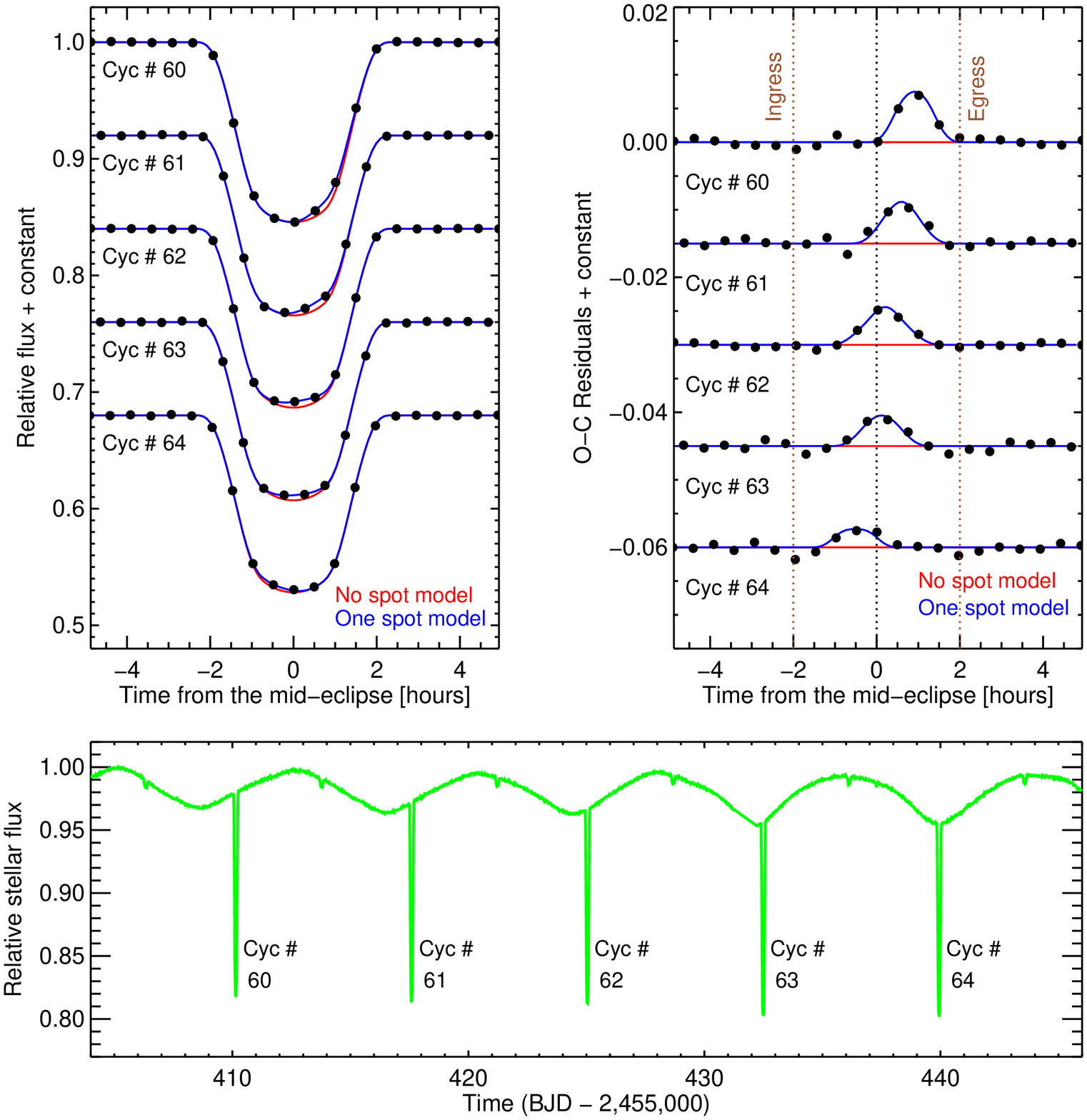}{oroszfig04.ps}{spot1}{\emph{Left:} 
Eclipses of a star spot in Kepler-47.  A dark spot on the
primary star rotates in and out of view, leading to the variable
flux in the out-of-eclipse regions as shown at the bottom.  
This star spot was partially
eclipsed by the secondary star during five consecutive primary
eclipses.  When the eclipse is modeled with a symmetric function,
a hump appears in the residuals.
As the spot moves on the primary,
the phase of the hump in the residuals
changes.  
\emph{Right:}
The O-C values of the primary eclipses in Kepler-47 vs the local
light curve slope.  The eclipses are systematically late when the
out-of-eclipse flux slopes downward during a primary eclipse
(this occurs when the dark
spot is rotating into view), and systematically
early when the out-of-eclipse flux slopes upward during a primary
eclipse (this occurs when the dark spot is rotating out of view).
Figures from Orosz et al.\ (2012). 
}

\articlefiguretwo{oroszfig05.ps}{oroszfig06.ps}{spot2}{Example 
CPOC diagrams where the ETVs are affected by
star spots, giving rise to ``random'' walk signals (the color
scheme is the same as the left panel in Figure
\protect{\ref{smallrms}}). From
top to bottom and left to right, 
the systems are KIC 4908495,
KIC 6697716, KIC 6706287, KIC 6863840,
KIC 7374746, KIC 9005854, KIC 10346522, and KIC 12418816.}

\subsection{EBs with Star Spots}

In an EB with immaculate stars, the eclipse profiles should
be smooth and symmetric in time for EBs with small eccentricities.
If a star spot (either dark or bright) appears on the star that
is being eclipsed, then the eclipse profile
will appear distorted.  
The distortion can be relatively large in cases
where the body in front is much smaller than the body 
that is being eclipsed.  If one is using a symmetric model
profile to find the time of minimum light, there will be a hump
in the residuals of the profile fit and the resulting
time will have a systematic error.  The primary eclipses
in Kepler-47 illustrate this quite nicely as shown in Figure 
\ref{spot1} (taken from Orosz et al.\ 2012, see also
Sanchis-Ojeda et al.\ 2012).
One or more dark spots on the primary rotate into and out of view,
leading to a modulation in the out-of-eclipse regions of
the light curve.  When the light curve has a downward slope
near primary eclipse, a dark spot is rotating into view,
and when the light curve has an upward slope near primary eclipse,
there is a dark spot rotating out of view.  Figure
\ref{spot1} shows a sequence of five primary eclipses where the local 
slope of the light curve goes from being positive 
to negative during the eclipse.  The spot is in
a different place on the primary star during 
successive primary eclipses, and
as a result the phase of the hump in the residuals changes.  
For Kepler-47 and many other EBs, there is a correlation between
the ETV of an eclipse time and the local slope of the light curve
during that eclipse (Sanchis-Ojeda et al.\ 2012), as shown
in Figure \ref{spot1}.  The presence of this correlation in Kepler-47
is a good indication that the signal seen in the ETVs of the primary
star is spurious, as the ETVs, corrected using this correlation, show
no signal (Orosz et al.\ 2012).

Because star spots seem to come and go, an EB with spots may exhibit
a ``random walk'' signal in the CPOC diagram. Such a signal could have systematic
deviations much larger than the nominal uncertainties in the
ETs, but will not remain coherent over the long term.
During the visual inspections of the CPOC diagrams, 117 EBs with
random walk signals were identified, and Figure \ref{spot2} shows
eight example cases.

\articlefiguretwo{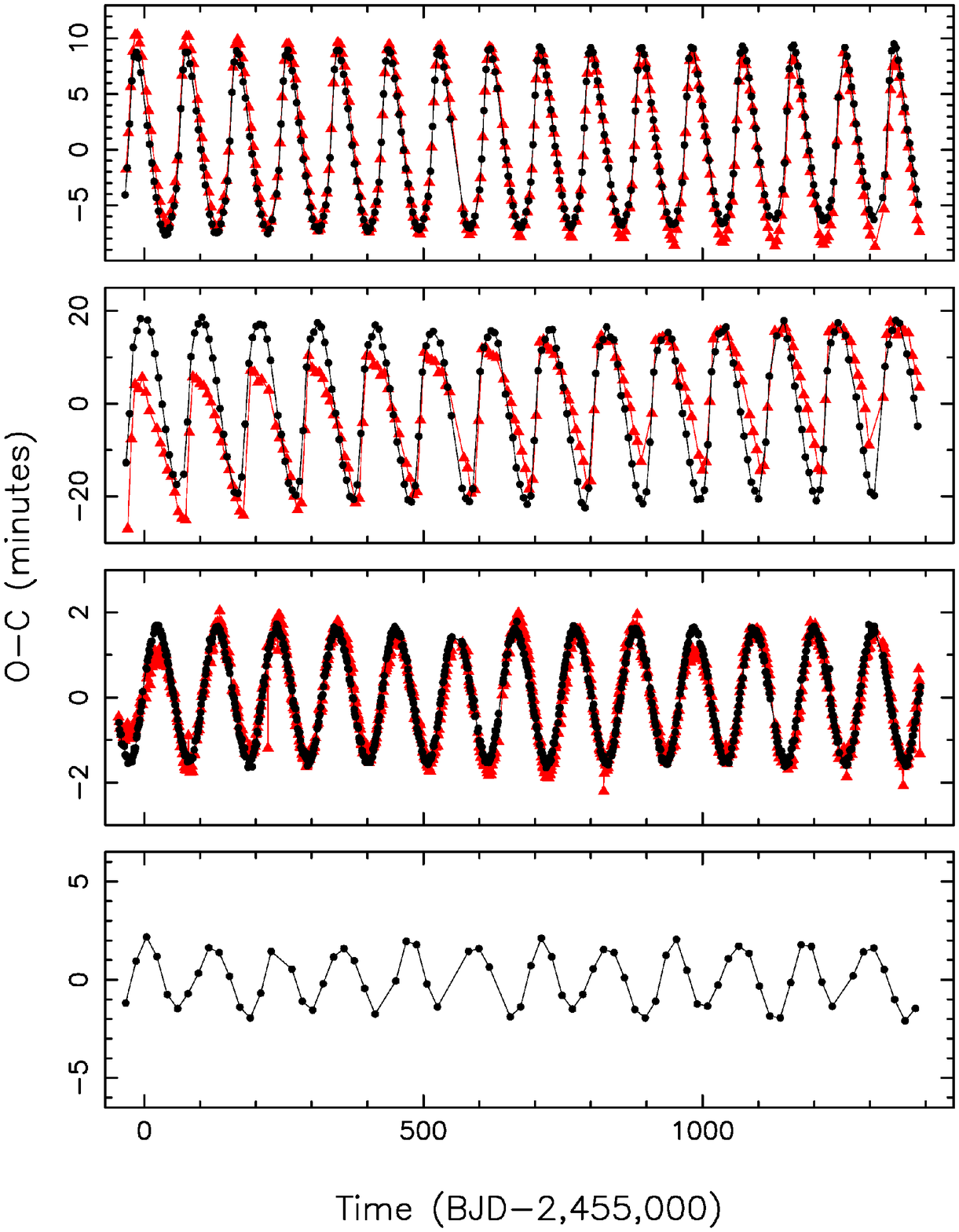}{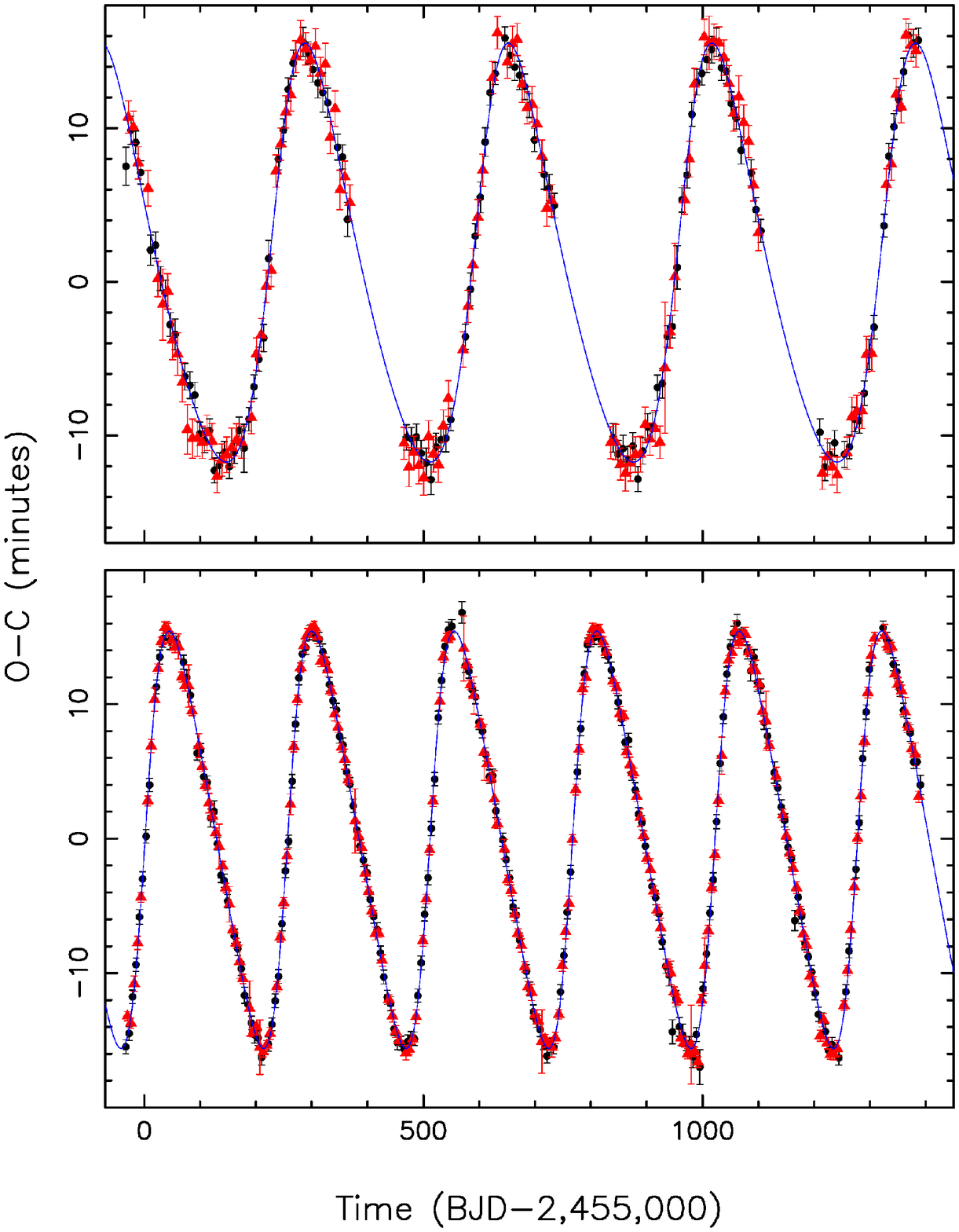}{periodic}
{\emph{Left:} Example CPOC diagrams for systems with obvious periodicities,
where the color scheme is the same as in Figure \protect{\ref{smallrms}}.
From top to bottom, the EBs are KIC 6545018 
($P_{\rm bin}=3.99$ d, $P_{\rm out}=90.4$ d),
KIC 9714358 ($P_{\rm bin}=6.47$ d, $P_{\rm out}=103.8$ d),
KIC 9451096 ($P_{\rm bin}=1.25$ d, $P_{\rm out}=106.9$ d), and
KIC 5095269 ($P_{\rm bin}=18.6$ d, $P_{\rm out}=117.9$ d).
\emph{Right:} Two EBs with periodic signals in their CPOC
diagrams where LTT model fits (blue lines)
yield very high tertiary masses.
From top to bottom, the EBs are 
KIC 4940201 ($P_{\rm bin}=8.82$ d, $P_{\rm out}=363.7$ d, $f(M)=4.53\,
M_{\odot}$), and
KIC 5384802 ($P_{\rm bin}=6.08$ d, $P_{\rm out}=255.2$ d, $f(M)=13.92\,
M_{\odot}$).}

\subsection{EBs with Obvious Periodicities in the CPOC}

During our visual inspections of the O-C and CPOC diagrams,
we identified 55 EBs with obvious periodicities in the primary
ETVs, including 7 with periods smaller than 200 days.  Figure
\ref{periodic} shows O-C and CPOC diagrams for the four
shortest-period ones, with ETV periods of 90.4,
103.8, 106.9, and 117.9 days.  These periods almost certainly
indicate the period of the third body.  A literature search
found mention of three EBs that were known before the launch of
{\em Kepler} with  third bodies having periods smaller 
than 200 days.  

The EBs where the primary and secondary ETVs are periodic and track each
other in the CPOC diagram can be modeled with an LTT orbit
(Equation 2) and the minimum mass (hereafter the 
``mass function'') of the third body can be computed
using Equation 4. A few EBs have implausibly large mass functions,
and Figure \ref{periodic} shows the CPOC diagrams of the
two EBs with the largest mass functions ($4.53\,M_{\odot}$
and $13.92\,M_{\odot}$).  As shown by Borkovits et al.\ (2003)
and Rappaport et al.\ (2013), dynamical effects can sometimes
produce ETV
signals  that can mimic ETVs due to pure LTT effects.  Thus
one should use extreme caution when modeling ETV signals with
a simple LTT model.

\articlefiguretwo{oroszfig09.ps}{oroszfig10.ps}
{plotlarge}{Example CPOC systems
where the range in the ETVs is too large to be 
caused by LTT effects only
(the color scheme is the same as in Figure \protect{\ref{smallrms}}).
From top to bottom and left to right, 
the systems are KIC 5255552 ($\Delta_{O-C}=13.3$
hr), KIC 7668648 ($\Delta_{O-C}=8.7$ hr),
KIC 5653126 ($\Delta_{O-C}=6.7$ hr), KIC 
7955301 ($\Delta_{O-C}=3.6$ hr),
KIC 7955301 ($\Delta_{O-C}=3.6$ hr), 
KIC 5771589 ($\Delta_{O-C}=3.5$ hr), 
KIC 5003117 ($\Delta_{O-C}=2.5$ hr), and
KIC 8210721 ($\Delta_{O-C}=2.5$ hr).
}

\subsection{EBs with Large ETVs, Changing Eclipse Depths, and
Tertiary Eclipses}

We have found about a dozen systems where the range on the ETVs
alone rules out simple LTT effects as the sole cause of the
variations.  Figure \ref{plotlarge} shows the eight systems
with the largest ETV spreads.  KIC 5255552 has a spread of
about 13.2 hours, which would require a displacement of the binary
by about 100 AU if the ETVs were due entirely to LTT effects.  We
conclude the cause of these ETV signals is largely dynamical.

There are at least 16 EBs that show significant changes in the
eclipse depths due to dynamical interactions.\footnote{There are 
also many cases where there are spurious changes in the eclipse
depths due to changing amounts of contamination due to nearby stars.
Different photometric apertures may be used for different
{\em Kepler} Quarters, and these different apertures can result
in different amounts of contamination.}
Figure \ref{depthchange} shows the light curves and CPOC
diagrams for eight of these depth-changing systems.
In all cases, there are significant ETVs, which is another 
indication that the depth changes are due to dynamical 
interactions.

Although hundreds of EBs with a tertiary
companion were known prior to the launch
of {\em Kepler}, none were observed to have eclipse events due to 
the third star.  KOI-126 was the first system with 
tertiary events identified in the {\em Kepler} data
(Carter et al.\ 2011).  We have identified 24 EBs that
show additional eclipse events due to a third body, including
the circumbinary planet systems (there are a few other cases
where there appear to be two unrelated EBs on the same
{\em Kepler} pixel). Figure \ref{tertfig}
shows eight examples.  The presence of these additional eclipse events
provides very strong constraints on the parameters of the system.
Modeling these light curves can be difficult, but one is then
rewarded with extremely precise parameters (e.g.\
Carter et al.\ 2011; Doyle et al.\ 2011).

\articlefigurefour{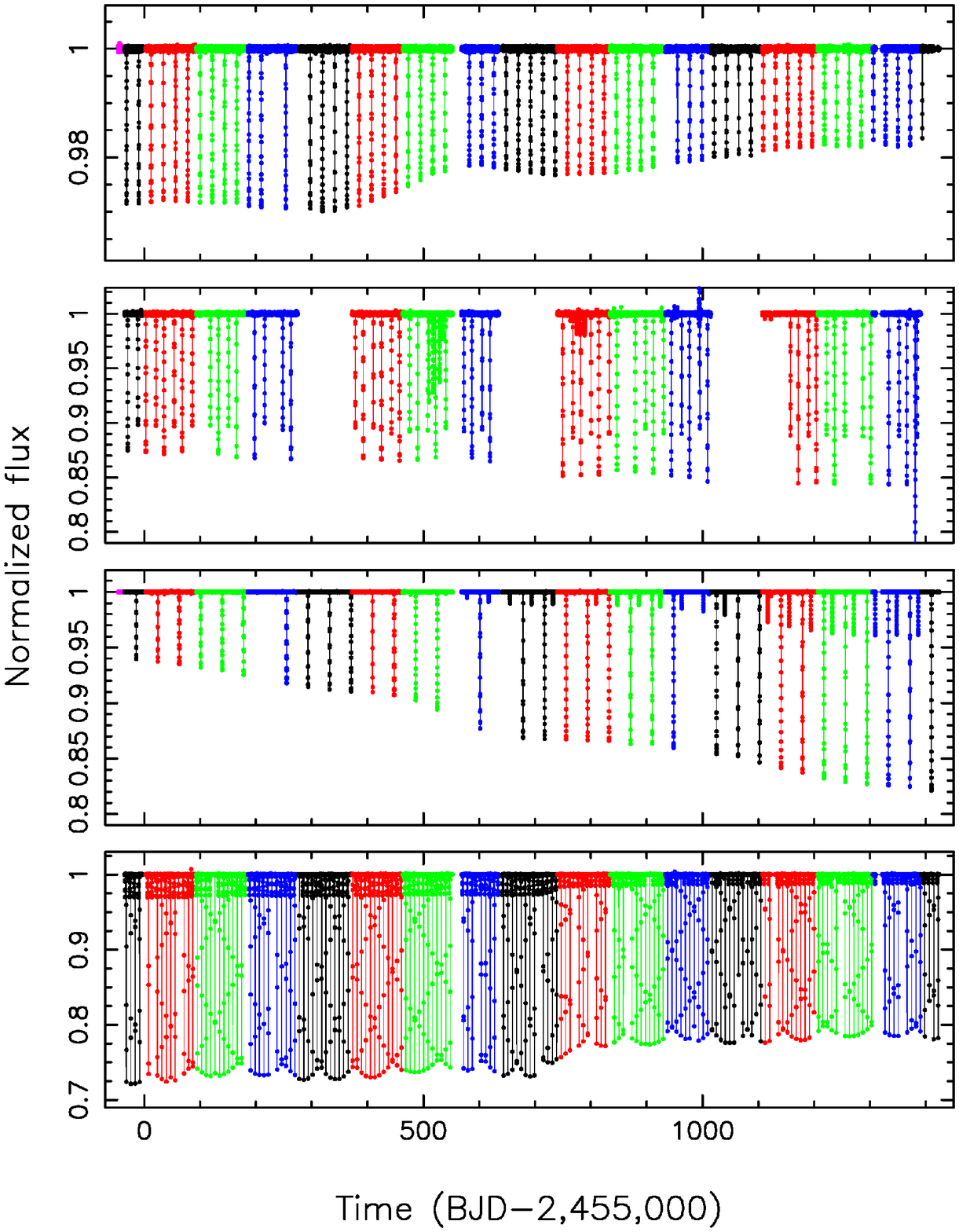}{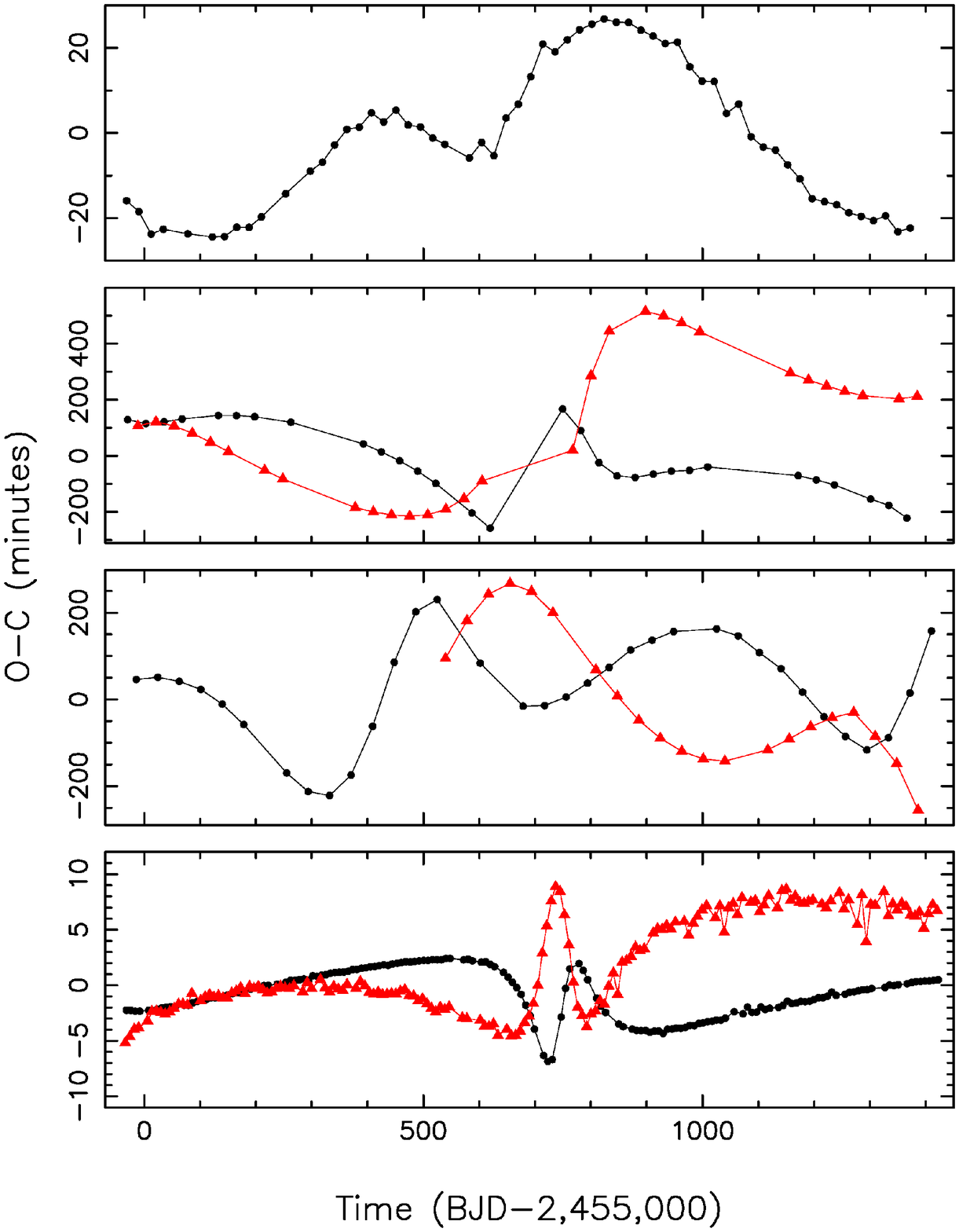}
{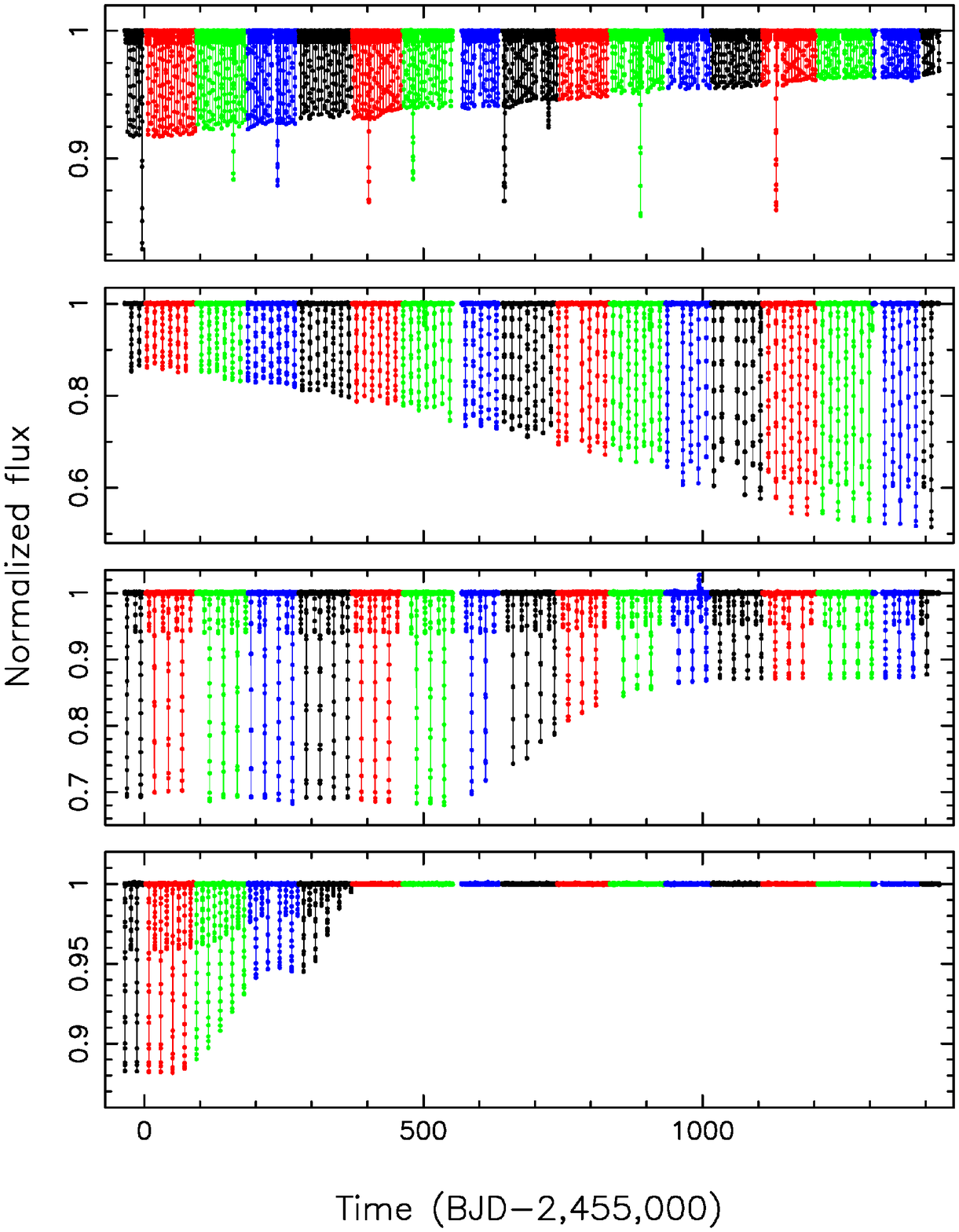}{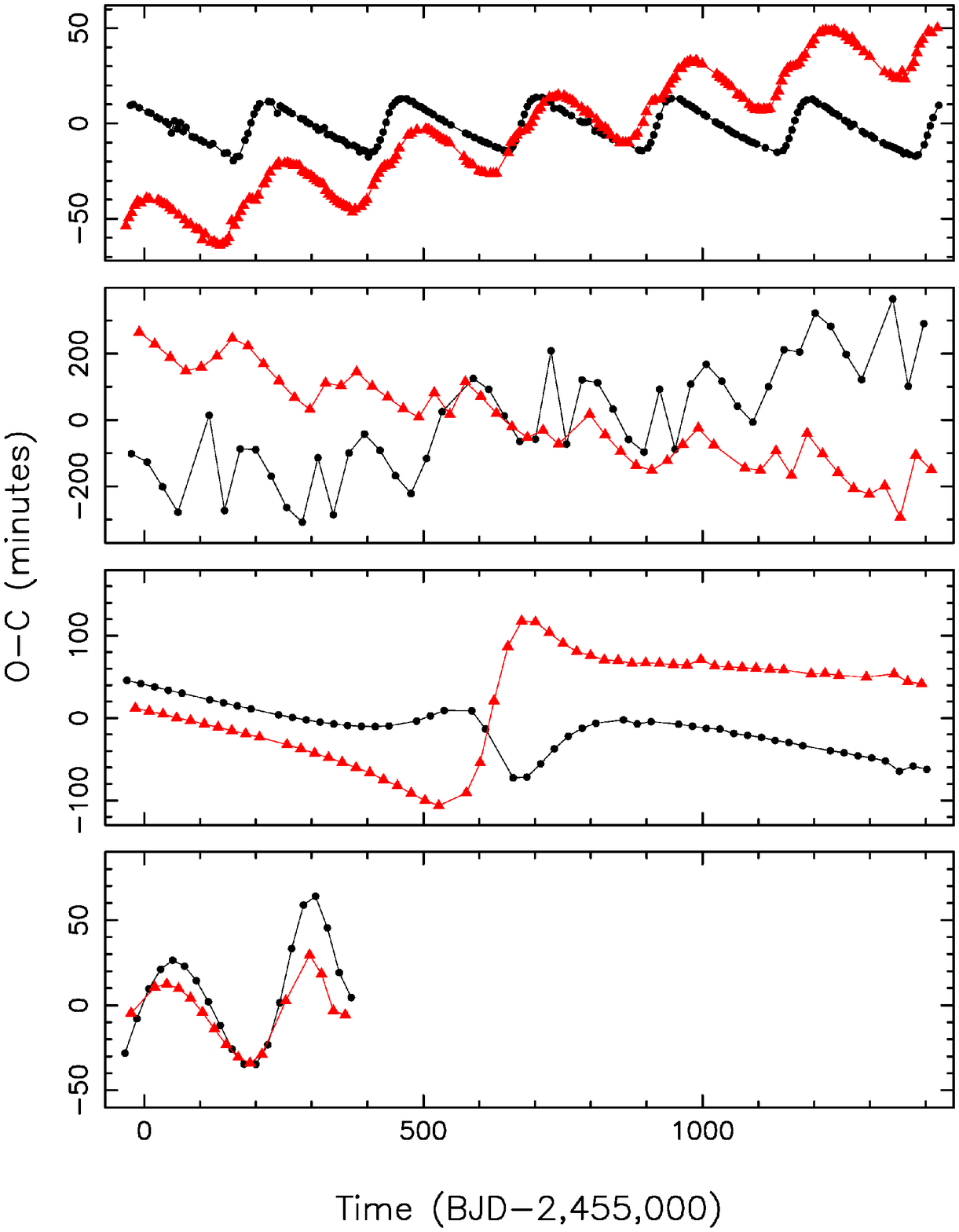}{depthchange}
{Example EBs where the depths of the eclipses have changed due
to the influence of a third body.  The normalized {\em Kepler}
light curves are shown on the left, where different colors
denote the {\em Kepler} Quarter (black for Q1, Q5, Q9, Q13, and Q17;
red for Q2, Q6, Q10, and Q14; green for Q3, Q7, Q11, and Q15;
and blue for Q4, Q8, Q12 and Q16).  The corresponding CPOC diagrams
are shown on the right with the same color scheme as in Figure 
\protect{\ref{smallrms}}.
From top to bottom, the systems are KIC 4769799,
KIC 5255552 (this system fell on the bad CCD module during 
Q5, Q9, Q13, and Q17), KIC 5653126
(note the appearance of the secondary eclipses starting in Q8),
KIC 5731312, KIC 7289157 (note the additional events due to the third body),
KIC 7668648, KIC 7670617, and KIC 10319590.  
}

\articlefiguretwo{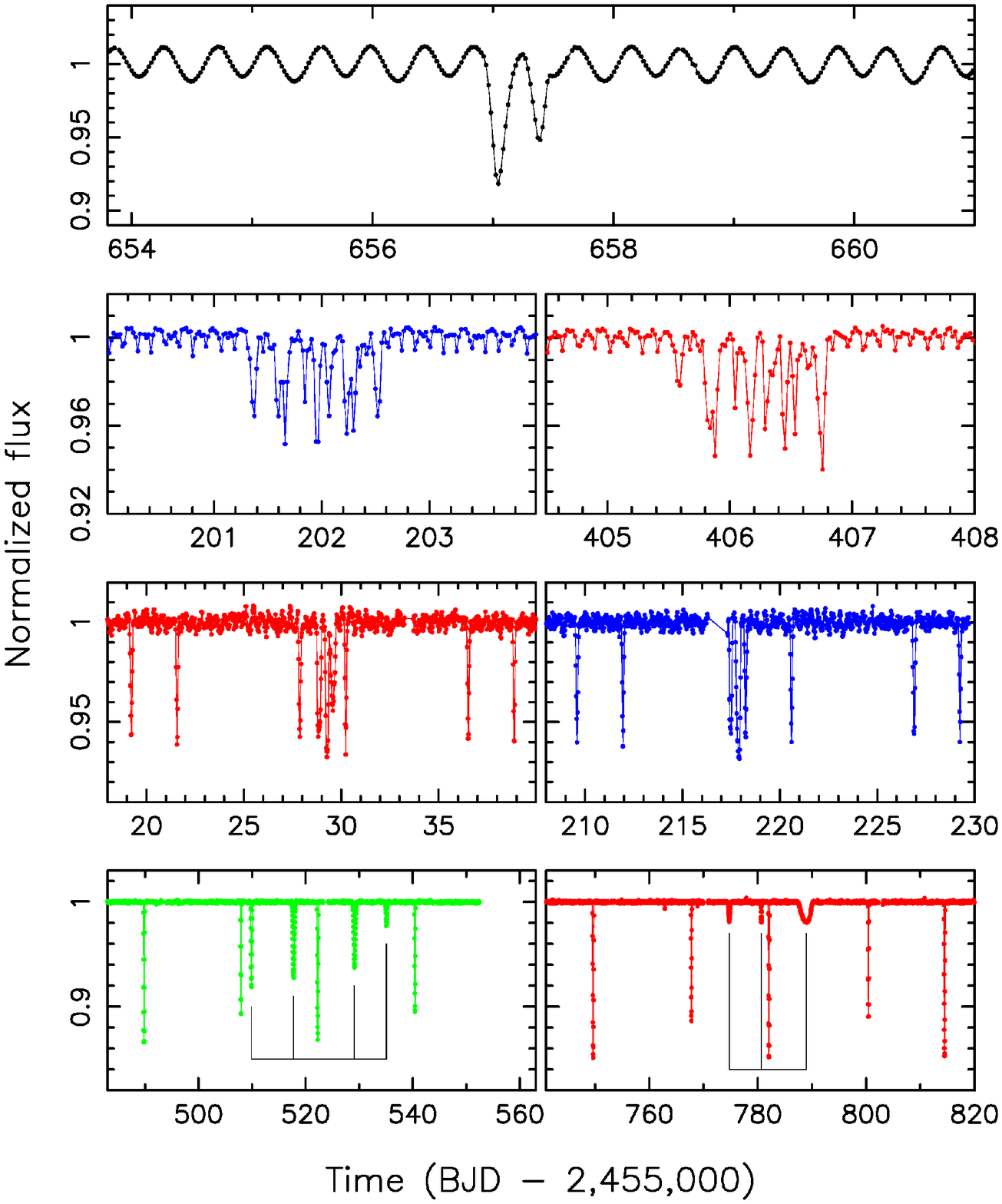}{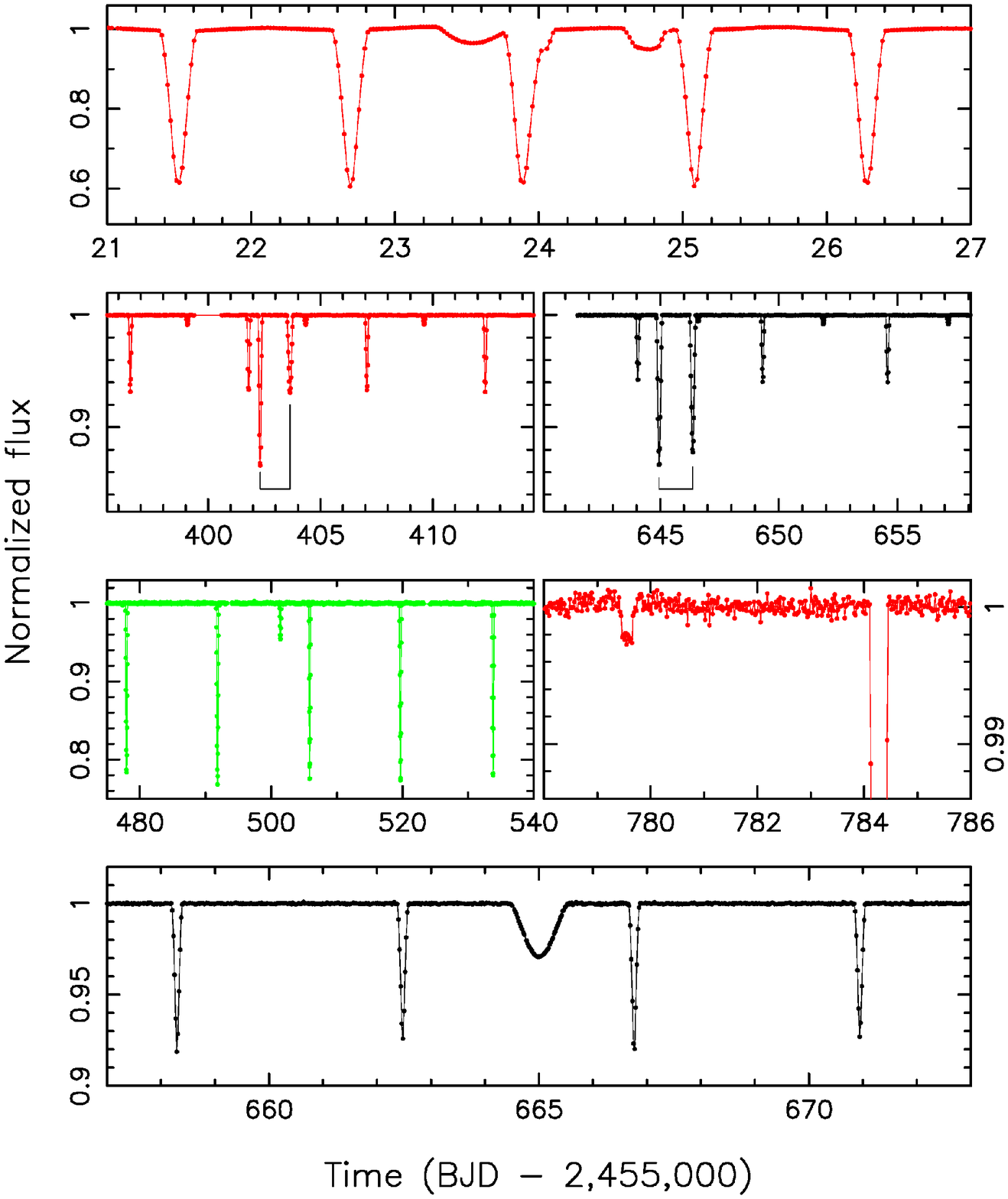}
{tertfig}{Some EBs that show tertiary events due to a third star.
The color scheme for the light curves is the same as in Figure
\protect{\ref{depthchange}}.
From left to right and top to bottom, the EBs are
KIC 2835289, KIC 2856960, KIC 4150611, KIC 5255552, KIC
6543674, KIC 7289157, KIC 7668648, and KIC 7670485.}

\subsection{EBs with Long-Term Trends or Diverging CPOCs}

We found 140 EBs with long-term trends in the ETVs.  These trends
presumably represent a small part of a signal with a periodicity
that is much longer than our $\approx 4$ year baseline.  Figure
\ref{trend} shows four examples.  The ETVs for the primary usually,
but not always, track the ETVs for the secondary in the CPOC diagram.

We also found
107 EBs where the primary ETVs and secondary ETVs have 
roughly linear trends with opposite 
slopes in the CPOC diagram.  
The ETVs with opposite slopes occur when only a small part of 
an apsidal period is observed.
Unlike the cases with large ETVs or periodic ETVs,
having opposite linear trends in the CPOC diagram does not necessarily
indicate the presence of the third body.  As noted earlier,
apsidal motion can be caused by tidal effects and/or General
Relativity.  While there are  good analytic approximations for
the rate of the apsidal advance, one needs
to know the masses to apply the expression for GR precession and 
the masses and fractional radii to apply the 
expression for the tidal apsidal
precession.  If one assumes masses near a solar mass for each
star, then detailed models
of the light curves can give good estimates of the eccentricity of the
binary and the fractional radii, making the expected rates 
of apsidal motion due to 
tides and GR computable.  We are in the process of 
modeling the light curves to determine what fraction of the EBs where 
the opposite trends in the CPOC diagram could be explained by GR, tidal
apsidal motion, or a combination of the two.    In the absence of detailed
light curve modeling, one can make a quick estimate by considering
only the systems with orbital periods longer than about 20 days, as the  
expected apsidal rates for GR and tides fall off rapidly with
increasing period, leaving the influence of a third body as the probable
cause of the opposite trends in the CPOC diagram.

\articlefiguretwo{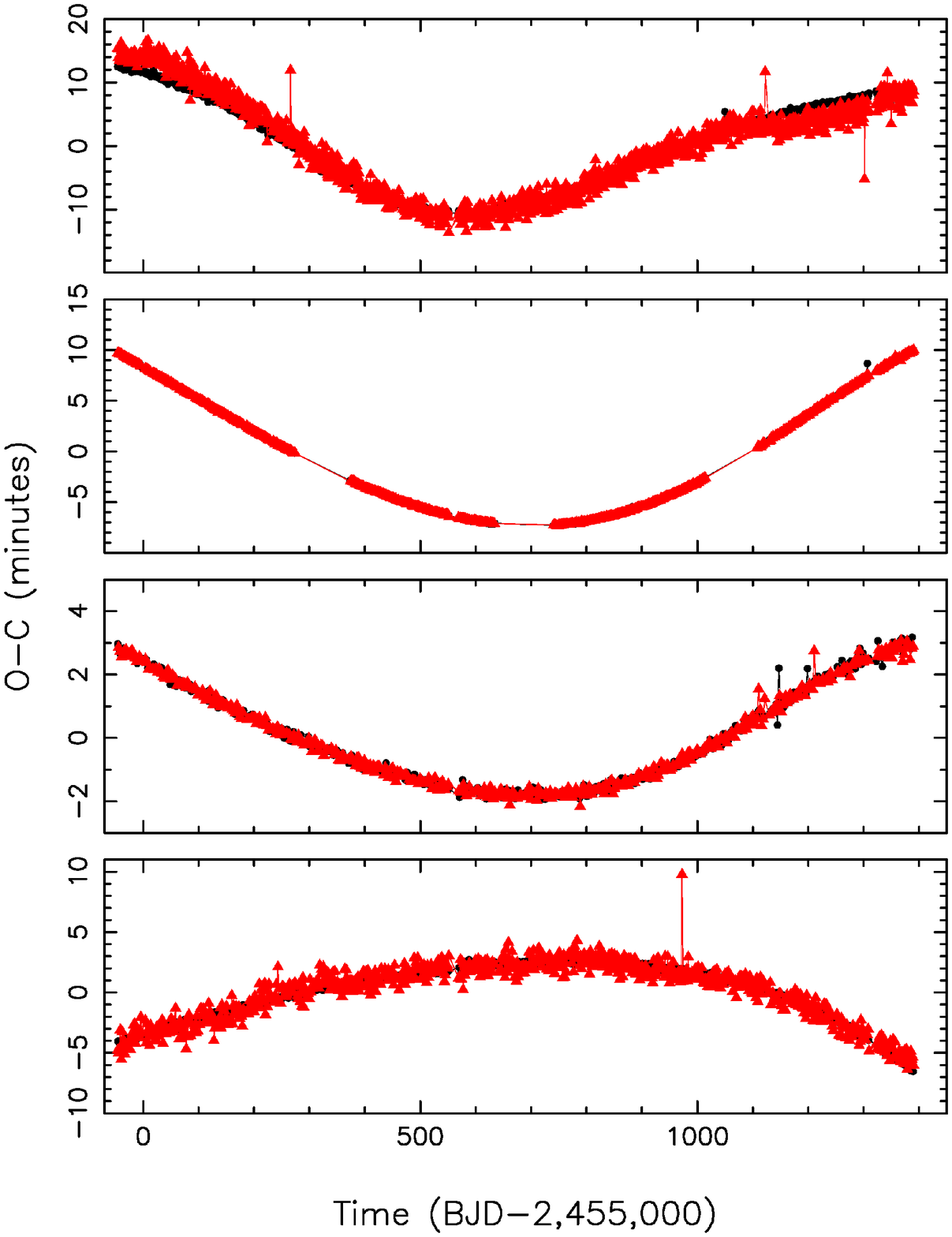}{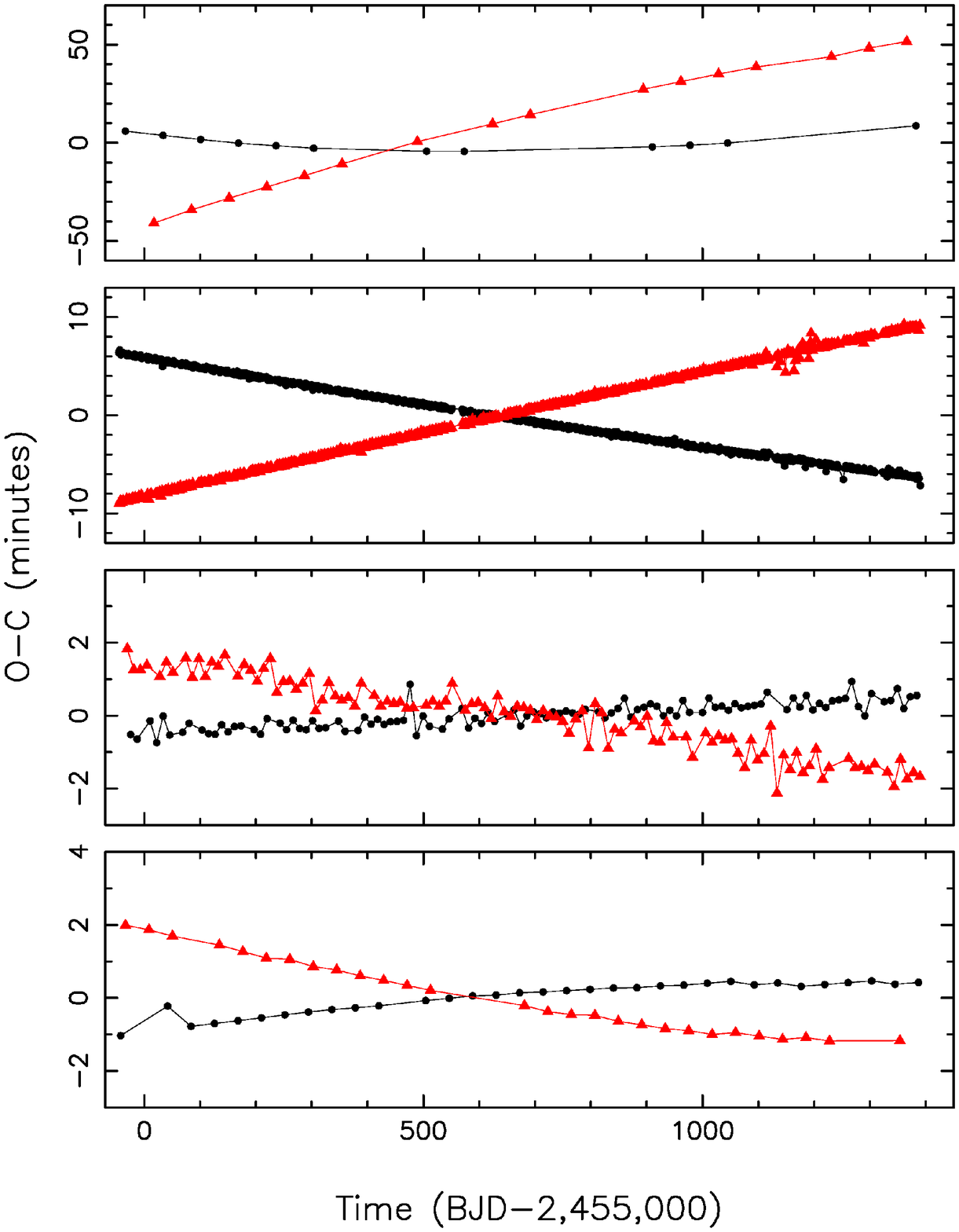}
{trend}{\emph{Left:}
Some EBs that show long-term trends in their ETVs.
From top to bottom, the EBs are
KIC 4732015, KIC 5513861, KIC 8429450, and KIC
8553788.  \emph{Right:}
Some EBs that have rougly linear trends with opposite
slopes in the CPOC diagram.
From top to bottom, the EBs are 
KIC 3247294, KIC 4544587, KIC 5955321, and KIC 8553907.}

On a related note, when the primary ETVs and the secondary ETVs show
opposite trends in the CPOC diagram, this means that one 
would get two different periods when fitting the primary ETs
and the secondary ETs separately.  We find 184 EBs where
the primary eclipse period differs from the secondary eclipse period
by more than $3\sigma$.  Not all of these cases have roughly linear
signals in the CPOC (for example KID 7289157 in Figure \ref{depthchange}),
and therefore
the number of systems found by this test exceeds the 107 systems found
above.  There are 136 EBs out of the 184 with orbital periods 20 days
or shorter, so in many of these cases the period differences could be
due to GR and/or tides.  For example, the primary and secondary
periods of KIC 4344587 differ by $\approx 901\sigma$ (the mean period
is about 2.19 days).  Gies et al.\ (2012) showed that this period
difference could be plausibly explained by tidal apsidal motion, as
both stars in this short-period eccentric binary have relatively large
fractional radii.  On the other hand, we have 48 systems with significant
period differences where the mean orbital period is longer than 20 days, and
the influence of a third body will almost certainly be needed to 
explain the period differences in these cases.

\subsection{A Brief Note on the Occurrence Rate of Close Triples}

In their limited survey of 41 EBs, Gies et al.\ (2012) found
14 EBs with long-term trends, which represents 34\% of the sample.
They argued that ``this finding is consistent with the presence of
tertiary companions among a significant fraction of the targets,
especially if many have orbits measured in decades''.  In their
more extensive survey, Rappaport et al.\ (2013) found 
39 candidate triple systems and after accounting for the types of
systems to which their search was sensitive, concluded that
at least 20\% of all close binaries have third body companions.
Conroy et al.\ (2014) found 236 candidate triple systems out
of 1279 searched, for a rate of about 18\%.

Although this work is ongoing, we can make some preliminary
estimates of the occurrence rate of triples in our sample.  The
O-C diagrams for the primary eclipses were visually inspected, and
the ETV signal was assigned one of 6 classes:  1 for flat
signal, 2 for an obviously periodic signal, 3 for a trend that
is concave upwards, 4 for a trend that is concave downward, 5 for
a ``wiggle'' (see KIC 5731312 in Figure \ref{depthchange}), and
6 for a random walk due to star spots.  The ETV signals for
categories 2, 3, 4, and 5 will most likely be due to third bodies.  
There were 203 systems with one of these four designations out of
the 1249 systems measured, which is about 16\%. If we use the 914 EBs with
acceptable profile fits as the sample size, then the fraction is 22\%.
In addition, the CPOC diagrams for all systems with both measured primary
and secondary eclipses were inspected, and the ones where the
primary ETV signal diverged from the secondary ETV signal were flagged.
There were 112 such cases.  In total there are 285 unique systems from
both lists, which is 22.8\% of the sample using a sample size of 1249 or
31\% using a sample size of 914.  
As explained earlier,
some of the EBs with roughly linear trends with opposite slopes in the CPOC
diagram may not have third bodies (e.g.\ apsidal precession due to GR or
tides may be sufficient to explain the ETVs).  Thus, the 
22.8\% or 31\% figures are upper limits.  Nevertheless, we can say that
based on our results so far, we can corroborate 
the results of Rappaport et al.\
(2013) who concluded that the occurrence rate of triples is on the order
of 20\%.

\subsection{Models of Individual Systems}

We have begun a program to systematically model the light and velocity
curves of several EBs with large ETVs and/or tertiary eclipse events.
The usual binary light curve synthesis codes are not adequate as
the positions of the stars are not described by simple Keplerian orbits.
We modified the ELC code (Orosz \& Hauschildt 2000) to include a dynamical
integrator that solves the Newtonian equations of motion to give the
positions of the stars at any given time.  Given the positions of
the stars, their radii, their relative flux contributions, and their
limb darkening properties, model light and velocity curves can be computed.

One of the more striking EBs is KIC 10319590, where the eclipses
disappeared after Q4 (see Figure \ref{depthchange}).  Spectra were
obtained using the echelle spectrograph on the Kitt Peak 4m telescope.
Lines from the primary and the tertiary star were detected.  Our best-fitting
to the light and velocity curves is shown in Figure \ref{disappear}.  
The three masses are $M_1=0.95\,M_{\odot}$,
$M_2=0.71\,M_{\odot}$, and $M_3=0.84\,M_{\odot}$.  The inner and outer periods
are 21.31 and 457.6 days, respectively.  
The outer orbit has a moderate eccentricity ($e=0.142$) and is
inclined by $43.1^{\circ}$ relative to the binary orbit.
This mutual inclination is consistent with the predictions of
the Kozai Cycle-Tidle Friction model for close binary formation
where one expects clusters of mutual inclinations near
40 and 140 degrees (Mazeh \& Shaham 1979;
Fabrycky \& Tremaine 2007).

\articlefigure[angle=-90, width=.72\textwidth]{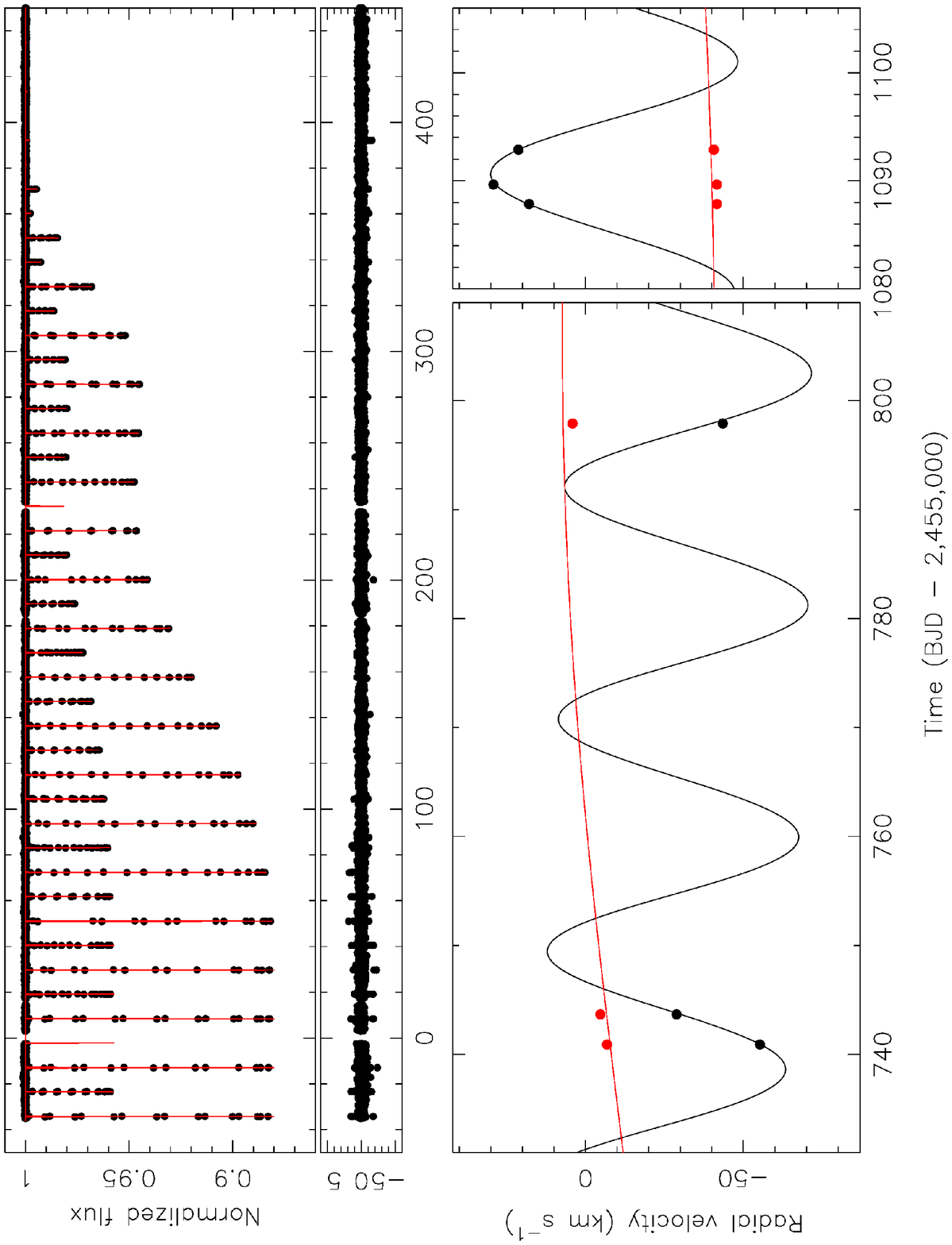}
{disappear}{\emph{Top:}
the normalized light curve of KIC 10319590 (points)
with the best-fitting photodynamical
model (red line).  The units of the residuals are parts per thousand.
\emph{Bottom:} The radial velocities of the primary
(black points) and the tertiary star (red points) and the best-fitting
model curves.}

KIC 7668648 shows a gradual increase in the eclipse depths and
large ETVs (Figure \ref{depthchange}).  In addition, the third star
transits both stars in the EB and is itself occulted by the other stars  
(Figure \ref{tertfig}).  Spectra were
obtained using the echelle spectrograph on the Kitt Peak 4m telescope.
The spectra are double-lined where both stars in the EB are detected.
A good photodynamical model was developed, and
we find masses of $M_1=0.849\,M_{\odot}$,
$M_2=0.808\,M_{\odot}$, and
$M_3=0.278\,M_{\odot}$, and  radii of
$R_1=1.010\,R_{\odot}$, 
$R_2=0.881\,R_{\odot}$,  and
$R_3=0.289\,R_{\odot}$.  The mean periods are 27.094 days for the
inner orbit and 206.4 days for the outer orbit.
Figure \ref{kid766} shows the best-fitting photodynamical model, and one
can see that  
the
transits and occultations  are well-fit, as are all of the primary
and secondary eclipses.  There is an odd feature in this
EB.
The eclipse event near day $-23$ (labeled $P_1$) is deeper than the 
eclipse event near
day $-9$ (labeled $S_1$), so the former was called a ``primary''
eclipse.  The eclipse event labeled $P_{52}$ occurs
51 orbital periods later than $P_1$ and the eclipse
event labeled $S_{51}$
likewise occurs 51 orbital periods later than $S_1$.  However,
$S_{52}$ is deeper than $P_{52}$, which means the roles of the primary
and secondary eclipses have reversed!  KIC 7668648 is a very important
system as it has a relatively low-mass star with a good mass
and radius determination and also from a dynamical point of view, as
it is a relatively compact triple.

\articlefiguretwo{oroszfig20.ps}{oroszfig21.ps}
{kid766}{\emph{Left:}
the normalized light curve of KIC 7668648 showing transits of the primary
and secondary by the third body with the best-fitting photodynamical
model (top) and occultations of the third
star by the primary and secondary (bottom).  
\emph{Right:} Close-up views of the first pair of primary and secondary eclipses
(top) and the last pair of primary and secondary eclipses (bottom).
Note that the 
roles of the primary and secondary eclipses have switched.
}





\acknowledgements It is a pleasure to thank the many collaborators
who have assisted with this work, including William Welsh,
Donald Short, Gur Windmiller, and numerous students at San Diego
State University, and members of the {\em Kepler} EB and TTV working
groups including Andrej Pr\v sa, Laurance Doyle, Robert Slawson,
Kyle Conroy, Steven Bloemen, Thomas Barclay,
Joshua Pepper, David Latham, Tsevi Mazeh, Joshua Winn,
Daniel Fabrycky, Joshua Carter, Jack Lissauer,
Eric Ford, Eric Agol, Darin Ragozzine,
Jason Steffen, Matt Holman, Billy Quarles, and Nader Haghighipour.
We acknowledge support from the U.S.\
National Science Foundation (grants AST-1109928 and AST-0850564) and
NASA (grant NNX12AD23G and the Kepler GO Program).






\end{document}